\documentclass{nature}

\usepackage{url}
\usepackage{amsmath}
\usepackage{amssymb}
\usepackage{color}
\usepackage[pdftex]{graphicx}
\usepackage{lineno}

\title{Coincidence of a high-fluence blazar outburst  with a PeV-energy neutrino event}

\author{M.\,Kadler$^{1}$\footnote{Corresponding Author}, F.\,Krau\ss$^{1,2}$, K.\,Mannheim$^1$, R.\,Ojha$^{3,4,5}$, C.\,M\"uller$^{1,6}$, R.\,Schulz$^{1,2}$, G.\,Anton$^{7}$, W.\,Baumgartner$^{3}$, T.\,Beuchert$^{1,2}$, S.\,Buson$^{8,9}$, B.\,Carpenter$^{5}$, T.\,Eberl$^{7}$, P.\,G.\,Edwards$^{10}$, D.\,Eisenacher~Glawion$^1$, D.\,Els\"asser$^1$, N.\,Gehrels$^3$, C.\,Gr\"afe$^{1,2}$, H.\,Hase$^{11}$, S.\,Horiuchi$^{12}$, C.\,W.\,James$^{7}$, A.\,Kappes$^{1}$, A.\,Kappes$^{7}$, U.\,Katz$^{7}$, A.\,Kreikenbohm$^{1,2}$, M.\,Kreter$^{1,7}$, I.\,Kreykenbohm$^{2}$, M.\,Langejahn$^{1,2}$, K.\,Leiter$^{1,2}$, E.\,Litzinger$^{1,2}$, F.\,Longo$^{13,13}$, J.\,E.\,J.\,Lovell$^{15}$, J.\,McEnery$^3$, C.\,Phillips$^{10}$, C.\,Pl\"otz$^{11}$, J.\,Quick$^{16}$, E.\,Ros$^{17,18,19}$, F.\,W.\,Stecker$^{3,20}$, T.\,Steinbring$^{1,2}$, J.\,Stevens$^{10}$, D.\,J.\,Thompson$^{3}$, J.\,Tr\"ustedt$^{1,2}$, A.\,K.\,Tzioumis$^{10}$, J.\,Wilms$^2$, J.\,A.\,Zensus$^{17}$}

\usepackage{ulem}
\begin{document}

\maketitle
\begin{affiliations}
 \item Institut f\"ur Theoretische Physik und Astrophysik, Universit\"at W\"urzburg, Emil-Fischer-Str. 31, 97074 W\"urzburg, Germany 
 \item Dr. Remeis Sternwarte \& ECAP, Universit\"at Erlangen-N\"urnberg, Sternwartstrasse 7, 96049 Bamberg, Germany
 \item NASA, Goddard Space Flight Center, Greenbelt MD 20771, USA
 \item University of Maryland, Baltimore County, Baltimore MD 21250, USA
 \item Catholic University of America, Washington DC 20064, USA 
 \item Department of Astrophysics/IMAPP, Radboud University Nijmegen, PO Box 9010, 6500 GL, Nijmegen, The Netherlands
 \item ECAP, Universit\"at Erlangen-N\"urnberg, Erwin-Rommel-Str. 1, 91058 Erlangen, Germany
 \item Istituto Nazionale di Fisica Nucleare, Sezione di Padova, 35131 Padova, Italy
 \item Dipartimento di Fisica e Astronomia “G. Galilei”, Universita di Padova, 35131 Padova, Italy 
 \item CSIRO Astronomy and Space Science, ATNF, PO Box 76, Epping NSW 1710, Australia
 \item Bundesamt für Kartographie und Geod\"asie, 93444 Bad K\"otzting, Germany
 \item CSIRO Astronomy and Space Science, Canberra Deep Space Communications Complex, PO Box 1035, Tuggeranong ACT 2901, Australia
 \item Dip. di Fisica, Universita’ di Trieste, Trieste, Italy
 \item INFN, Trieste, Italy
 \item School of Mathematics \& Physics, University of Tasmania, Private Bag 37, Hobart, 7001 Tasmania, Australia
 \item Hartebeesthoek Radio Astronomy Observatory, Krugersdorp, South Africa
 \item Max-Planck-Institut f\"ur Radioastronomie, Auf dem H\"ugel 69, 53121 Bonn, Germany
 \item Observatori Astron\`omic, Universitat de Val\`encia, C/ Catedr\'atico Jos\'e Beltr\'an no. 2, 46980 Paterna, Val\`encia, Spain 
 \item Departament d'Astronomia i Astrof\'isica, Universitat de Val\`encia, C/ Dr. Moliner 50, 46100 Burjassot, Val\`encia, Spain
 \item Department of Physics and Astronomy, University of California at Los Angeles, Los Angeles, CA 90095
\end{affiliations}

\begin{abstract}
The discovery of extraterrestrial very-high-energy neutrinos by the IceCube collaboration has launched a quest for the identification of their astrophysical sources.
Gamma-ray blazars have been predicted to yield a cumulative neutrino signal exceeding the atmospheric background
above energies of 100 TeV, assuming that both the neutrinos and the $\gamma$-ray photons are produced by accelerated protons in 
relativistic jets.
Since the background spectrum falls steeply with increasing energy, the individual events with the clearest signature of being of an extraterrestrial origin are those at PeV energies.
Inside the large positional-uncertainty fields of the first two PeV neutrinos detected by IceCube,
the integrated emission of the blazar population   
has a sufficiently high electromagnetic flux to explain the detected IceCube events, but fluences of individual objects are too low to make an unambiguous source association.
Here, we report that a major outburst of the blazar PKS~B1424$-$418 occurred in temporal and positional coincidence with the third PeV-energy neutrino event (IC\,35) detected by IceCube.
Based on an analysis of the full sample of $\gamma$-ray blazars in the IC\,35 field and assuming a photo-hadronic emission model,
we show that the long-term average $\gamma$-ray emission of blazars as a class is in agreement with both the measured all-sky flux of PeV neutrinos and the spectral slope of the IceCube signal.
The outburst of PKS~B1424$-$418 has provided an energy output high enough to explain the observed PeV event, indicative of a direct physical association.
\end{abstract}

The neutrino excess detected by IceCube comprises 37 events with energies between 30\,TeV and 2\,PeV, rejecting a purely atmospheric origin at a significance of 5.7 standard deviations\cite{Aar13,Ice13,Ice14}.
These events show a broad distribution across both hemispheres of the sky consistent with an extragalactic source population.  
Due to the very steep background of atmospheric neutrinos, events at PeV energies are best suited for attempting to establish associations with individual blazars.
In the first two years of observations, IceCube detected two events with about 1\,PeV of deposited energy\cite{Aar13,Ice13} (IC\,14, and IC\,20; dubbed `Bert' and `Ernie'). A third event
at 2\,PeV (IC\,35; dubbed `BigBird') was recorded in the third year of IceCube data\cite{Ice14} on 2012 Dec 4. The IceCube analysis concentrated on very-high-energy events with interaction signatures that were fully contained within the detector (High Energy Starting Events; HESE).
In combination with an equal-neutrino-flavor flux at Earth\cite{Iceflavor}, this resulted in the majority of the detected events being cascade-like, with relatively large median positional uncertainties ($R_{50}$) of typically $10^\circ$ to $20^\circ$.
While a number of different source classes have been discussed as a possible origin of a diffuse neutrino flux\cite{Man92,Man95a,Ste13,Fox13,Tay14,Pad14,Kra14,Mur14,Bec14}, no individual astrophysical object has been identified so far from which a neutrino flux with a substantial Poisson probability for a detection by IceCube is expected.

Due to the Earth's opacity, the IceCube HESE analysis detects events at PeV energies mainly\footnote{A different analysis of IceCube muon neutrinos finds an excess signal also from the northern sky\cite{Ice15a}.} from the southern sky\cite{Ice14}.
Thus,
contemporaneous astronomical data to probe the various source hypotheses can best be obtained via southern-hemisphere monitoring programs. 
TANAMI\footnote{\url{http://pulsar.sternwarte.uni-erlangen.de/tanami}} is a multiwavelength program\cite{Ojh10,Kad15} that monitors the brightest $\gamma$-ray-loud active galactic nuclei (AGN) located in the southern sky at declinations below $-30^\circ$. It comprises the ideal database to estimate the diffuse neutrino flux due to the integrated
emission of AGN in a given large field at a given time, as well as the maximum possible neutrino flux associated with an individual object of the sample.  

Blazars are radio-loud AGN with jets oriented close to the line of sight. This substantially increases the apparent brightness of these objects owing to the Doppler boosting of the emission from the relativistically moving emission zones.
A direct association of a PeV-neutrino with an individual $\gamma$-ray blazar would have the important implication that a sizeable fraction of their observed $\gamma$-ray emission must be due to hadronic decays, and that blazar jets are also sources of ultra-high-energy cosmic rays\cite{Hil84}.
The X-ray and $\gamma$-ray emission of blazars may originate from the photoproduction of pions by accelerated protons\cite{Man89}. 
Protons that are accelerated in the jet (e.g., via shock acceleration) could interact with `seed' photons (e.g., ultraviolet photons from the accretion disk surrounding the central supermassive black hole). The resulting cascades produce charged and neutral pions, which decay and produce neutrinos and high-energy photons. Simple estimates and detailed Monte Carlo simulations show\cite{Mue00,Kra14} that in this scenario $F_\gamma \lesssim F_\nu$,
where the X-ray to $\gamma$-ray flux is 
integrated over the high-energy
spectral energy distribution (SED). If the seed photons are provided by a blue/UV bump component, as is typical in the blazar subclass of flat spectrum radio quasars (FSRQs), the neutrino spectrum is expected to peak at PeV energies\cite{Kra14}.
Attributing the high-energy electromagnetic emission to these photohadronic processes, the maximum possible neutrino PeV emission can be estimated from the measured integrated flux of high-energy photons.

Using TANAMI multiwavelength data, we previously compiled and discussed the multiwavelength properties of the six radio- and $\gamma$-ray-brightest blazars located inside the $R_{50}$ fields of the two $\sim 1$\,PeV events IC\,14 and IC\,20 from the first two years of IceCube data\cite{Kra14}.
 We found  relatively low maximum neutrino fluxes of these six individual blazars owing to their low fluence built-up over two years, but the
diffuse flux due to the integrated emission of all blazars in the fields was found to be sufficiently high to expect up to two events. 
When the contribution of the large number of fainter sources from the blazar population is taken into account\cite{Kra15}, the maximum possible neutrino flux inside a given field is increased further.
A high-angular-resolution point-source search with the ANTARES neutrino telescope found
a signal flux fitted by the likelihood analysis corresponding to approximately one event for each of the two blazars with the highest predicted neutrino fluxes, Swift~J1656.3$-$3302 and TXS~1714$-$336. This result is 
consistent with the blazar-origin hypothesis of IC\,14 but it is also consistent with the hypothesis of a background signal\cite{A+T15a}. 
No events were found for the IC\,20 candidate blazars, constraining the range of possible neutrino spectra to spectral indices flatter than $-2.4$ for the blazar-origin scenario. While no conclusive association could be found, this result demonstrates the potential of identifying individual neutrino blazar sources, if suitable high-fluence candidates can be found.

\begin{figure}
\centering
\includegraphics[width=0.9\linewidth]{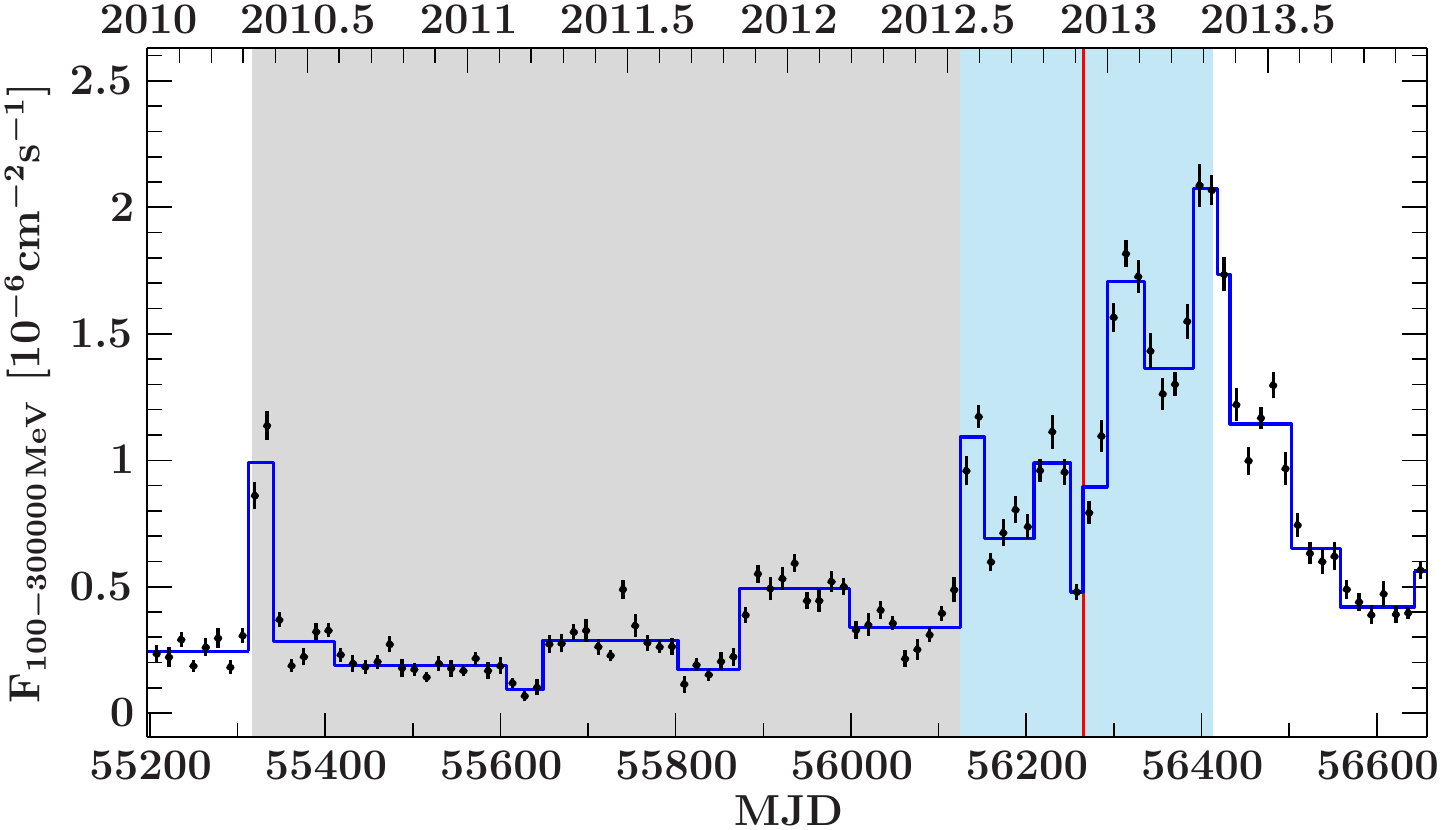}
\hspace*{0.015\linewidth}\includegraphics[width=0.875\linewidth]{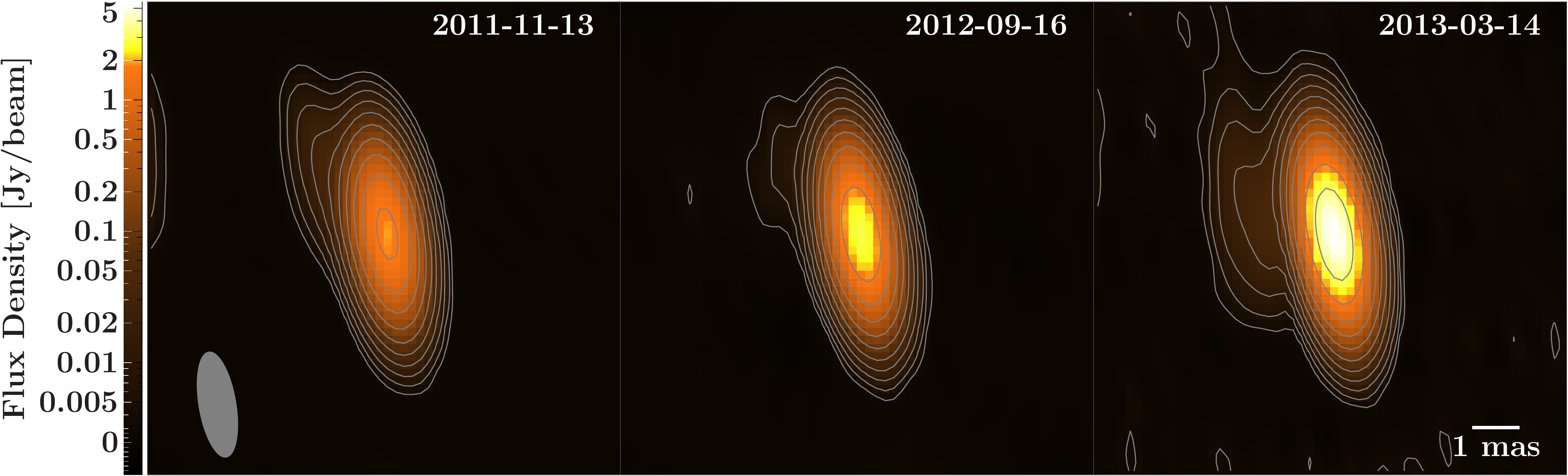}
\caption{\small \textbf{a) $\gamma$-ray light curve of PKS~B1424$-$418}. The \textit{Fermi}/LAT data are shown as two-week binned photon fluxes between 100\,MeV and 300\,GeV (black), the Bayesian blocks light curve (blue), and the IC\,35 time stamp (red line). The first three years of IceCube integration (2010 May through 2013 May) and the included outburst time range are highlighted in color.
\label{fig:fig1}
\textbf{b) TANAMI VLBI images of PKS~B1424$-$418}. The images show the core region at 8.4\,GHz from 2011 Nov, 2012 Sep and 2013 Mar in uniform color scale. 1\,mas corresponds to about 8.3\,pc. All contours start at $3.3$\,mJy\,beam$^{-1}$ and increase logarithmically by factors of 2. The images were convolved with the enclosing beam from all three observations of $2.26\textrm{\,mas} \times 0.79\textrm{\,mas}$ at a position angle of $9.5^\circ$, which is shown in the bottom left. The peak flux density increases from 1.95\,Jy\,beam$^{-1}$ (2011 Apr) to 5.62\,Jy\,beam$^{-1}$ (2013 Mar).
\label{fig:fig2}
}
\end{figure}

\textbf{Coincidence of a PeV neutrino with a major blazar outburst}\\
The third PeV neutrino (IC\,35) detected by the IceCube collaboration\cite{Ice14} had an energy of $2004^{+236}_{-262}$\,TeV and a median positional uncertainty of $R_{50}=15.9^\circ$ centered around the coordinates $\mathrm{RA}=208.4^\circ$, $\mathrm{Dec}=-55.8^\circ$ (J2000). Following our earlier strategy\cite{Kra14}, we searched this field for positional coincidences with $\gamma$-ray-emitting AGN. 
In the second catalog of AGN detected by the \textit{Fermi} Large Area Telescope (2LAC)\cite{2lac,lat}, which was based on \textit{Fermi}/LAT all-sky observations between 2008 Aug and 2010 Sep, a total of 20 $\gamma$-ray-bright AGN were found in the $R_{50}$ field of IC\,35. Seventeen of these AGN are blazars (2 FSRQs, 2 BL\,Lac objects, and 13 AGN of uncertain type), two are radio galaxies (Centaurus\,A and Centaurus\,B), and one is a starburst galaxy (NGC\,4945).
The radio galaxy Cen\,A is the closest AGN and the brightest radio source in the field. However, the bulk of the radio emission is emitted from the kpc-scale lobes of this FR\,I-type radio galaxy,
and Padovani \& Resconi (2014)\cite{Pad14} discard Cen\,A as a possible source of the IceCube event because the extrapolated SED at PeV energies is too low.
The dominant blazar in the field is PKS~B1424$-$418
at redshift $z=1.522$\cite{Whi88} and classified as an FSRQ. Due to its relatively low $\gamma$-ray flux in the first three months of the \textit{Fermi} mission in 2008, it was not included in the \textit{Fermi}/LAT Bright Source List\cite{Abdo09}. The source showed two $\gamma$-ray flares in 2009-2011\cite{Bus14} and is listed as a bright $\gamma$-ray source in all subsequent \textit{Fermi}/LAT catalogs. 
Still, Padovani \& Resconi (2014)\cite{Pad14} discarded it from their list of most-probable counterparts for the 2\,PeV IceCube neutrino due to its relatively low $\gamma$-ray emission in the 2008-2011 period.
In summer 2012, PKS~B1424$-$418 commenced a dramatic rise in $\gamma$-ray brightness\cite{Ojh12}.
In contrast to previous flares, this increase marked the beginning of a long-lasting high-fluence outburst over more than a year with $\gamma$-ray fluxes exceeding 15 to 30 times the flux reported in the \textit{Fermi} 2LAC (see Fig.~\ref{fig:fig1}) and which coincides with the PeV neutrino event IC\,35 both in position and in time. 
With a $\gamma$-ray photon fluence of $(30.5 \pm 0.3)$\,cm$^{-2}$, PKS~B1424$-$418 \textsl{showed the absolute highest 100\,MeV to 300\,GeV $\gamma$-ray fluence of all extragalactic sources} in the 2012 Jul 16 through 2013 Apr 30 period, which spans the arrival time of the PeV neutrino.

Along with the very bright $\gamma$-ray emission, an increase in X-ray, optical, and radio emission from PKS~B1424$-$418 has also been reported\cite{Cip13,Has13,Nem13}. 
Figure~\ref{fig:fig2} shows a series of TANAMI VLBI images of PKS~B1424$-$418 at 8.4\,GHz observed between 2011 Nov and 2013 Mar (see Supplementary `Methods').  They show that the sharp increase in radio flux density from $\sim 1.5$\,Jy to $\sim 6$\,Jy took place inside the VLBI core, i.e.,
on projected scales smaller than $\sim 3$\,pc (see cf. Supplementary Data Table~\ref{tab:vlbi-images}). 
The 2012 Sep image is the first VLBI epoch within the GeV high-fluence phase and also the first to show a substantial increase in the core flux density.
This high-amplitude radio outburst is unparalleled in the TANAMI sample since the beginning of the program in 2007.
A physical association of the outburst of PKS~B1424$-$418 and the PeV neutrino event is suggested given the
unprecedented nature of these two events and the small 
\textsl{a posteriori} probability for a chance coincidence of about 5\% (see Supplementary: `Methods').

\textbf{The maximum-possible number of PeV neutrinos from blazars in the IC\,35 field}\\
Can the calorimetric output of this single blazar outburst account for the necessary PeV neutrino flux associated with the IC\,35 event? And is a
possible association in agreement with the observed all-sky rate of PeV events and with the lack of obvious additional associations
of PeV events with other bright blazars? 
In the following, we 
use the IC\,35 field as  representative of the full sky to predict
the number of 
PeV neutrino events in IceCube from blazars, first as a population and second, as individual sources  
in which photomeson production gives rise to bright neutrino emission.
We then compare the predicted numbers to the observed IceCube results:
starting by considering electron neutrinos, which are prone to produce cascade events in the IceCube detector,
the maximum possible number of neutrinos detected in a solid angle $\Omega$ is (neglecting neutrino oscillations)
\begin{equation}
N_{\nu,\mathrm{PeV}}^\mathrm{max}(\Omega) = A_{\mathrm{eff},\nu_e}\cdot \left(\frac{F_\gamma}{E_\nu}\right) \cdot \Delta t \mathrm{\quad ,}
\end{equation}
where $F_\gamma$ is the $\gamma$-ray energy flux of all blazars located inside $\Omega$ integrated between 5\,keV and 10\,GeV,
$\Delta t = 988\mathrm{\,days}$ is the IceCube three year observation time, and
$A_{\textrm{eff},\nu_e}$ is the 
effective area of the IceCube HESE analysis\cite{Ice13} at PeV energies
for charged-current (CC) interactions of electron neutrinos.
So far, 
three PeV neutrinos have been detected by IceCube, of which two had an energy of $\sim 1$\,PeV and one had an energy of $\sim 2$\,PeV. The IceCube effective area evaluated at the geometric mean of the three events' effective areas is 
$\sim 2.2 \cdot 10^5$\,cm$^{2}$.
The integrated emission from the 17 2LAC blazars in the field predicts a maximum of $\sim 7.9$ neutrino PeV events for 988\,days of IceCube integration (see Supplementary Data Table~\ref{tab:events}) but we also need to consider the contribution of fainter blazars, which are not listed as resolved sources in the 2LAC catalog. 
In total, blazars make up $\sim 50$\,\% of the extragalactic $\gamma$-ray background EGB\cite{diffusegamma,ajello15} 
but the integrated flux for all 2LAC blazars inside $\Omega_\mathrm{IC\,35}$ is only $F_{100\mathrm{\,MeV}-820\mathrm{\,GeV}} = 8.5 \cdot 10^{-7}$\,ph\,cm$^{-2}$\,s$^{-1}$. Distributed over $\Omega_\mathrm{IC\,35}$, this corresponds to $3.5 \cdot 10^{-6}$\,ph\,cm$^{-2}$\,s$^{-1}$\,sr$^{-1}$, which accounts for only about 30\,\% of the EGB\footnote{Given the rather shallow slope of the $\log N \log S$ distribution\cite{3lac}, future \textit{Fermi}/LAT catalogs will mitigate this effect only slowly.} 
Thus, 
we may expect about $\frac{0.2}{0.3} \cdot 7.9 \sim 5.3$ additional neutrinos at PeV energies
from faint unresolved blazars within $\Omega_\mathrm{IC\,35}$ (taking their EGB contribution as a proxy for their integrated keV-to-GeV output). Thus, the maximum possible number of neutrinos predicted by this model from blazars in the IC\,35 field is
\begin{equation}
N_{\nu,\mathrm{PeV}}^\mathrm{max}(\Omega_\mathrm{IC\,35}) \sim 13 \quad ,
\end{equation}
which includes maximum-possible PeV neutrino counts from all $\gamma$-ray blazars from the 2LAC catalog plus a maximum-possible contribution of the large population of faint unresolved blazars.

\textbf{Predicted all-sky count of PeV neutrinos from blazars and comparison to observation}\\
By extrapolating from the fairly representative field $\Omega_\mathrm{IC\,35}$,
we estimate the maximum number of PeV neutrino events from all
blazars (both resolved and unresolved) over three years from the full southern sky to be
\begin{equation}
N_{\nu,\mathrm{\,PeV}}^\mathrm{max}(2\pi) = 13 \cdot \frac{2\pi}{\Omega_\mathrm{IC\,35}} \sim 336 \quad .
\end{equation}
This number of events would be expected 
\textsc{i)} if only electron neutrinos would be produced,
\textsc{ii)} if all blazars harbored dense UV photon fields due to the emission of optically thick accretion disks as is typical for FSRQs, and
\textsc{iii)} if the neutrino spectrum peaked sharply at PeV energies.
All three conditions are  clearly not fulfilled as only 3 events have been detected,  leading to an empirical scaling factor of
\begin{equation}
f_\mathrm{emp}=\frac{N_{\nu,\mathrm{\,PeV}}^\mathrm{obs}(2\pi)}{N_{\nu,\mathrm{\,PeV}}^\mathrm{max}(2\pi)} \sim \frac{3}{336} \sim 0.009 \quad .
\label{eq:f_emp}
\end{equation}
This can be compared to a theoretical value $f_\mathrm{th}$, which 
accounts for physically motivated realistic deviations from the three ideal conditions. The theoretical scaling factor allows us to
predict the number of detectable PeV events $N_{\nu}^\mathrm{pred}$ as
\begin{equation}
N_{\nu,\mathrm{PeV}}^\mathrm{pred}(\Omega_\mathrm{IC\,35}) = f_\mathrm{th} \cdot N_{\nu,\mathrm{PeV}}^\mathrm{max}(\Omega_\mathrm{IC\,35}) \quad . 
\label{eq:f}
\end{equation}
The scaling factor is factorized into a flavor factor $f_\mathrm{\textsc{i}}$, a factor accounting for the different classes of blazars $f_\mathrm{\textsc{ii}}$, and a spectrum factor $f_\mathrm{\textsc{iii}}$: 
\begin{equation}
f_\mathrm{th} = f_\mathrm{\textsc{i}} \cdot f_\mathrm{\textsc{ii}} \cdot f_\mathrm{\textsc{iii}} \quad \mathrm{.}
\end{equation}
The IceCube data indicate an equal flavor ratio\cite{Iceflavor} so that
the flavor factor would be $1/3$ if 
only electron neutrinos are accounted for when computing the maximum event numbers.
When adding the two other flavors, it has to be considered that 
the number of detected cascade events due to muon and tau neutrinos is lower than for electron neutrinos because of the energy-dependent cross sections and inelasticities for neutral-current (NC) and charged-current interactions.
Assuming an underlying neutrino power law with slope $-2.3$, as observed by IceCube\cite{Ice13}, we estimate a fraction of $f_\mathrm{\textsc{i}} \sim 0.5$ for cascade events at $(1-2)$\,PeV in IceCube.
The deepest available \textit{Fermi}/LAT point source catalogs contain a fraction of FSRQs of about $f_\mathrm{\textsc{ii}} \sim 0.5$ (and about the same numbers of BL\,Lac objects)\cite{2lac}. 
For our basic model of a sharply peaked neutrino spectrum due to photopion production from monoenergetic UV photons, $f_\mathrm{\textsc{iii}}$ would be equal to unity.
In a more realistic scenario, 
a range of Doppler shifts (depending on the location of the seed-photon sources with respect to the relativistic jet base as discussed in our earlier work\cite{Kra14}) causes broader spectra extending to lower neutrino energies. Considering also broadening due to the different redshifts of sources, an output range of \mbox{$\sim 30$\,TeV} to \mbox{$\sim 10$\,PeV} can be expected. 
In addition, models which consider 
proton-proton collisions or assume accretion tori with virial temperatures of $\sim 10^9$\,K rather than optically thick accretion disks\cite{Man93,Man95a,Man95b} also
predict softer spectra.
Using a spectral index of $-2.3$ as measured by IceCube\cite{Ice14} and the \mbox{($30$\,TeV} to \mbox{$10$\,PeV)} bandwidth of the spectrum reduces the number of PeV output neutrinos by $f_\mathrm{\textsc{iii}} = 0.05$, so that we estimate
\begin{equation}
f_\mathrm{th} = 0.5 \cdot 0.5 \cdot 0.05 \sim 0.0125 
\end{equation}
(cf. Eq.~\ref{eq:f_emp}).
Our model thus predicts $0.0125 \cdot 336 \sim 4$ events at PeV energies from the full southern sky, which is remarkably close to the observed three PeV events.
We conclude that the measured $\gamma$-ray emission of the blazars in the IC\,35 field allows us to reproduce both the measured all-sky flux of PeV neutrinos
and the measured spectral slope of the IceCube signal
assuming a simple photo-hadronic emission model of FSRQs.

\textbf{Predicted number of PeV neutrinos for individual FSRQs}\\
If $\Omega$ becomes small, containing only one individual FSRQ, we can set $f_\mathrm{\textsc{ii}} = 1$. The predicted number of PeV neutrinos for an individual FSRQ is then 
\begin{equation}
N_{\nu,\mathrm{PeV}}^\mathrm{pred}(\mathrm{FSRQ}) = 0.025 \cdot N_{\nu,\mathrm{PeV}}^\mathrm{max}(\mathrm{FSRQ}) \quad \mathrm{,}
\label{eq:fsrq}
\end{equation}
from which Poisson probabilities for detections of neutrinos from individual sources can be calculated.
For the 
2LAC sources in the IC\,35 field, we find relatively low maximum-possible neutrino values ($N_{\nu,\mathrm{PeV}}^\mathrm{max} \sim 0.04 - 0.9$) in 16 of the 17 cases,
from which small predicted neutrino counts are predicted ($N_{\nu,\mathrm{PeV}}^\mathrm{pred} \sim 0.001 - 0.023$),
corresponding to small individual Poisson probabilities for any neutrino detections during the 3-year IceCube integration of ($P \lesssim 0.1\,\% - 2.2\,\%$). 
PKS~B1424$-$418 in its pre-outburst state reached a maximal possible neutrino event number of $N_{\nu,\mathrm{PeV}}^\mathrm{max} \sim 1.6$ ($N_{\nu,\mathrm{PeV}}^\mathrm{pred}\sim 0.04, P \lesssim 3.9\,\%$). 

\begin{figure}
\centering
\includegraphics[width=0.9\textwidth]{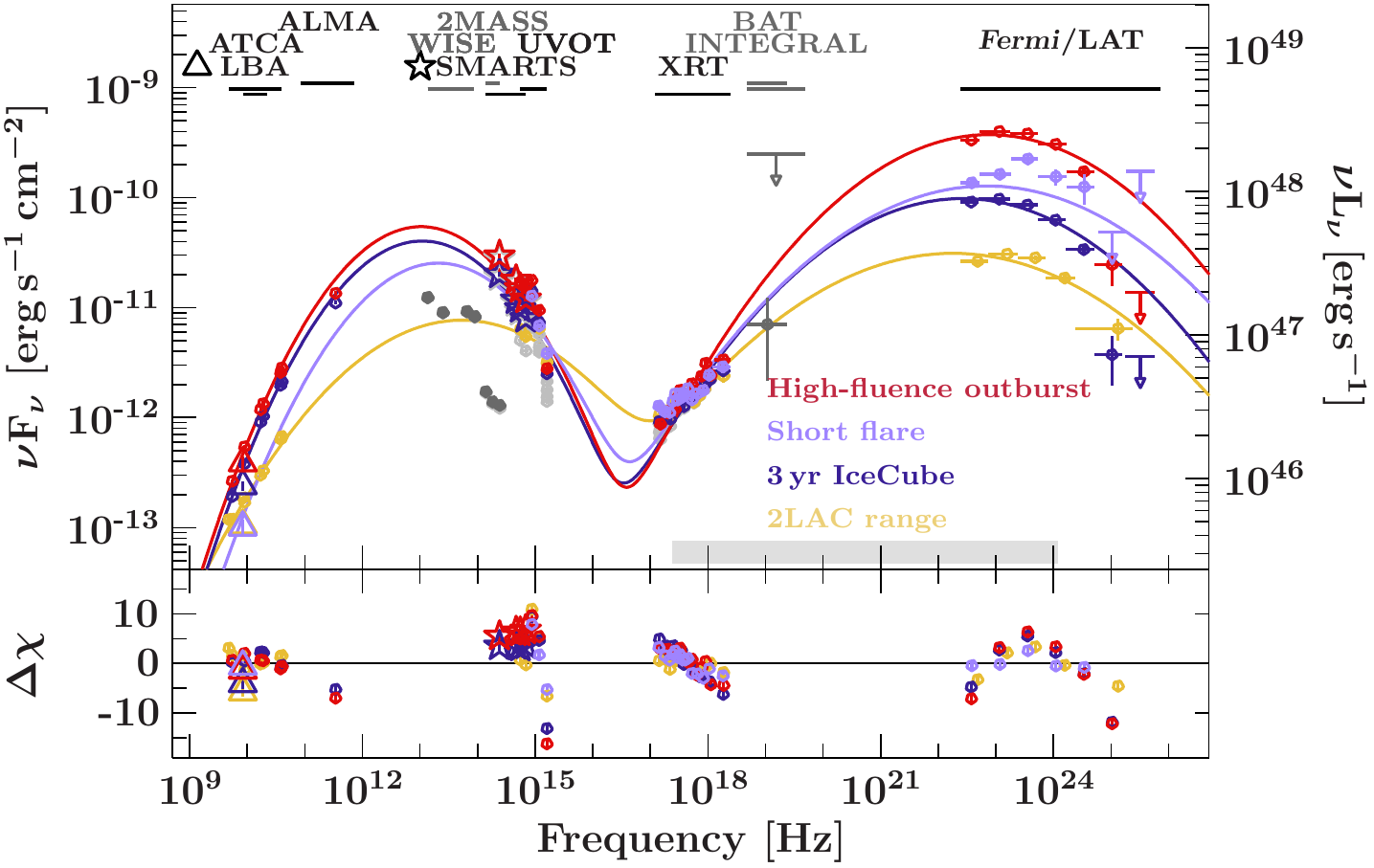}
\caption{\small \textbf{Dynamic SED of PKS~B1424$-$418}.
The multi-epoch SEDs are fitted with two log parabolas
for the 2LAC period (yellow), the first three years of IceCube integration (dark blue), the short 2010 flare around MJD\,55327 (red), and the major outburst between  MJD\,56125 and MJD\,56413 (light purple). The gray shaded area shows the keV to GeV integration range for the neutrino fluence calculation. Note that upper limits (downward arrows) are neglected in the fits. 
\label{fig:fig3}
}
\end{figure}

\textbf{Predicted number of PeV neutrinos for the PKS~B1424$-$418 outburst}\\
In Fig.~\ref{fig:fig3}, we show the average broadband SED of PKS~B1424$-$418 for the 2LAC period, the three year IceCube integration period, the 2010 short flare around MJD\,55327 (cf.~Fig.~\ref{fig:fig1}), and the major outburst phase between 2012 Jul 16 and the end of the IceCube period in 2013 Apr (see Supplementary: `Methods'  for details of the SED production). 
With the peak in the GeV range for all periods considered, and the X-ray spectrum changing mostly in photon index, we see that the \textit{Fermi}/LAT flux values shown in the light curve are good proxies for the integrated high-energy output.
In spite of the relatively high fluxes during the 2010 flare, the short duration yields only a small fluence, resulting in a low maximum-possible neutrino value of $N_{\nu,\mathrm{PeV}}^\mathrm{max} \sim 0.2$. Substantially higher values
are derived for the 2LAC period ($N_{\nu,\mathrm{PeV}}^\mathrm{max} \sim 1.6$, $N_{\nu,\mathrm{PeV}}^\mathrm{pred} \sim 0.04$) and the three year IceCube period ($N_{\nu,\mathrm{PeV}}^\mathrm{max} \sim 4.5$, $N_{\nu,\mathrm{PeV}}^\mathrm{pred} \sim 0.11$).
The latter is dominated by the 9-month period between the start of the outburst (2012 Jul 16) and the end of the three year IceCube period (2013 Apr 30). During these 9 months, the source increased its predicted neutrino-production rate by more than an order of magnitude 
so that the Poisson probability to detect a neutrino associated with the 9-month high-fluence outburst of PKS~B1424$-$418 is at a considerable level of about 11\,\%, which is three times higher than the corresponding probability to detect an event from the integrated emission of all other known $\gamma$-ray blazars in the field during this 9-month period. 
Our model thus allows us to associate an individual blazar during a rare major outburst with the highest-energy extraterrestrial neutrino detected by IceCube to date.
 
\textbf{Why don't we detect PeV neutrinos from every bright blazar?}\\
If our model is correct, it also has to explain the non-detection of PeV neutrinos in positional agreement with other high-fluence blazars and with the detection statistics of sub-PeV neutrino events. We note that the positional uncertainties $R_{50}$ given by the IceCube team are median values, which means that only half of all events originate inside their measured $R_{50}$ regions while the other half are coming from larger offset angles. Above, we have calculated the \textsl{maximum number of neutrino events that can be explained} by individual astrophysical sources within $R_{50}$ for a high-confidence event. When asking for the \textsl{maximum number of IceCube events that might be} associated with a given astrophysical source, a larger radius has to be considered. For example, within $2 \times R_{50}$, PKS~B1424$-$418 is in positional agreement with the sub-PeV events IC\,16 ($30.6^{+3.6}_{-3.5}$\,TeV at an offset of $1.5 R_{50}$) and IC\,25 ($33.5^{+4.9}_{-5.0}$\,TeV, $1.4 R_{50}$) so that the data are not in disagreement with a rather broad and steep neutrino spectrum. A point source search with ANTARES, following the strategy applied to the candidate blazars in the IC\,14 and IC\,20 fields\cite{A+T15a}, will be able to exclude the doublet/triplet hypothesis in case of a non-detection of any events. A preliminary analysis of the ANTARES collaboration\cite{A+T15b}finds no excess signal at the position of PKS~B1424$-$418, excluding the possibility of a very steep neutrino spectrum associated with the blazar outburst.

We have used the \textit{Fermi}/LAT monitored source list light curves\footnote{\url{http://fermi.gsfc.nasa.gov/ssc/data/access/lat/msl_lc/}} to identify candidate sources for high keV-to-GeV fluence, compiled the average SEDs over the three years of IceCube integration for the top-ten candidate sources from the whole sky and derived their expected neutrino counts (see Table~\ref{tab:rankedlist}). For northern- and southern-hemisphere events, we have used the effective areas for the appropriate minimum energy provided by the IceCube team\cite{Ice13}. We do not extend the list beyond rank 10, because for the tenth-ranked source, the maximum possible neutrino output has already dropped by more than an order of magnitude to ${\mathcal O}(1)$.
Only three other sources reach a predicted neutrino output comparable to PKS~B1424$-$418. The two FSRQs PKS~B1510$-$089 and 3C\,454.3 both have a maximum PeV-neutrino output on the order of $8$ in three years of IceCube integration but do not coincide with any of the three observed PeV events. Applying the source scaling factor of $0.025$ for the $\gamma$-ray blazars, the Poisson probability for detecting zero PeV events from a source of this fluence is $\sim 80$\,\%. 
On the other hand, the model predicts a $\sim 50$\,\% probability  to detect at least one neutrino from one of the four top-ranked high-fluence blazars and the detection of more than one PeV event remains at
a realistic probability of about 16\,\%. In this context, it is intriguing that  
also the gravitationally-lensed blazar PKS~B1830$-$211, which is the highest-ranked source in the top-10 blazar-fluence list, is located only marginally outside the $R_{50}$ region of the PeV event IC\,14, which was  detected by IceCube on 2011, Aug 09, coinciding with a 
high-fluence outburst phase of this blazar\footnote{\url{http://fermi.gsfc.nasa.gov/FTP/glast/data/lat/catalogs/asp/current/lightcurves/PKS1830-211_604800.png}}.
In addition, PKS~B1830$-$211 is positionally coincident with the $R_{50}$ fields of six additional sub-PeV IceCube neutrino events and with the region of the highest, albeit not significant, point-source clustering test statistic of IceCube events\cite{Ice14}. The list of coincidences includes the high-energy events IC\,2 ($117^{+15}_{-15}$\,TeV, $0.3 R_{50}$) and IC\,22 ($220^{+21}_{-24}$\,TeV, $1.2 R_{50}$). However,  ANTARES  has measured an upper limit of $1.89 \cdot 10^{-8}$\,GeV\,cm$^{-2}$\,s$^{-1}$ on the energy flux of neutrinos from PKS~B1830$-$211\cite{antares_lensedblazars}, assuming an $E^{-2}$ neutrino spectrum. This value is very similar to the limit found for PMN~J1802$-$3940, based on which an association with three or more IceCube neutrinos could be excluded  at 90\,\% confidence for neutrino spectral indices steeper than $-1.8$. 
The positional proximity of high-fluence blazars in our list to other IceCube sub-PeV events or even the temporal proximity to high-fluence phases (see Supplementary `Methods') is likely coincidental in most cases because the atmospheric contribution increases and the IceCube effective area decreases rapidly below 100\,TeV.

\textbf{Constraining the neutrino velocity}\\
Recently, a
theoretical limit of $(v - c)/c \le (0.5 - 1.0) \times 10^{-20}$ for superluminal neutrinos has
been derived from constraints on vacuum pair emission and neutrino splitting\cite{Ste14}.
Assuming a physical association between the outburst activity of PKS~B1424$-$418 and the IC\,35 PeV neutrino, an observational constraint on the neutrino velocity is implied:
the maximum possible time-travel delay between the beginning of the outburst and the arrival of the neutrino is $\sim 160$\,days, constraining the relative velocity difference to $(v-c)/c \lesssim {\mathcal O}(10^{-11})$ (for a light travel time of $9.12$ billion years). This is about two orders of magnitude more constraining than the neutrino-velocity limit derived from SN\,1987A\cite{Lon87}.
However, as discussed in the Supplementary `Methods', a $\sim 5$\,\% probability for a chance coincidence remains. It also cannot be excluded that the observed PeV neutrino could be associated with an historical (or future) 
outburst of the source.

\textbf{Summary and outlook}\\
Tentative associations of high-energy neutrinos with flaring blazars have been suggested before\cite{Hal05,ANTARES_Flare}
but it remained questionable whether a high-enough neutrino flux could be produced in the candidate flares\cite{Rei05}.
Here, we have identified for the first time a single source that has
emitted a sufficiently high fluence during a
major outburst to explain an observed coinciding PeV neutrino event.
There is a remarkable
coincidence with the IceCube-detected PeV neutrino
event IC\,35 with a probability of only $\sim 5$\,\% for a chance coincidence.
Our model reproduces the measured rate of PeV events detected over the whole sky by IceCube and 
accounts for the distribution of neutrino events accross the bandwidth expected for photo-hadronic
neutrino production. 
A substantial increase of the significance of putative future coincidences between PeV neutrino events 
and high-fluence blazars could be achieved considering  track events at smaller median angular errors or the observation of doublet events associated with the same blazar. 
However, it has to be kept in mind that only a small fraction of the total $\gamma$-ray emission of all blazars is associated with the brightest individual objects. In fact, only $\sim 70$\,\% of the blazar $\gamma$-ray emission has been resolved into point sources so far\cite{diffusegamma} by \textit{Fermi}/LAT. 
For any individual PeV neutrino event, there will thus always remain a large probability of being associated with the population of faint remote sources, which are not contained in the bright-source $\gamma$-ray catalogs. 
We thus expect three out of ten future PeV neutrinos to not be associated with any known $\gamma$-ray blazar\footnote{The recently reported multi-PeV neutrino-induced muon event
detected by IceCube\cite{Ice15b}, which does not coincide with any known bright $\gamma$-ray source, might well be an event of this type.}. 
Within the next years of IceCube observations, the combination of improved number statistics and 
continuous multiwavelength monitoring of high-fluence blazars
is the key 
to developing a consistent scenario of hadronic processes
in AGN jets and their long-suspected
association with extragalactic cosmic rays\cite{Hil84}.



\begin{addendum}
 \item[Acknowledgements] We acknowledge support and partial funding by the Deutsche
  Forschungsgemeinschaft grant WI 1860-10/1 (TANAMI) and GRK 1147,
  Deutsches Zentrum f\"ur Luft- und Raumfahrt grant
  50\,OR\,1311/50\,OR\,1303/50\,OR\,1401, 
  the German Ministry for Education and Research (BMBF) grants 05A11WEA, 05A14WE3
  the Helmholtz Alliance for Astroparticle Physics (HAP),
  the Spanish MINECO project AYA2012-38491-C02-01, the Generalitat
  Valenciana project PROMETEOII/2014/057, the COST MP0905
  action ``Black Holes in a Violent Universe''
  as well as NASA through Fermi Guest Investigator grants NNH09ZDA001N, NNH10ZDA001N, NNH12ZDA001N, and NNH13ZDA001N.
This study made use of data collected by the Australian Long Baseline Array (LBA)
and the AuScope initiative.
  The LBA is part of the Australia
Telescope National Facility which is funded by the Commonwealth of
Australia for operation as a National Facility managed by CSIRO.
AuScope Ltd is funded under the National Collaborative Research Infrastructure Strategy (NCRIS), an Australian Commonwealth Government Programme.
This paper has also made use of up-to-date SMARTS optical/near-infrared light curves that are available at www.astro.yale.edu/smarts/glast/home.php.
The \textit{Fermi}-LAT Collaboration acknowledges support for LAT
development, operation and data analysis from NASA and DOE (United
States), CEA/Irfu and IN2P3/CNRS (France), ASI and INFN (Italy), MEXT,
KEK, and JAXA (Japan), and the K.A.~Wallenberg Foundation, the Swedish
Research Council and the National Space Board (Sweden). Science
analysis support in the operations phase from INAF (Italy) and CNES
(France) is also gratefully acknowledged.
We thank J. E. Davis and T. Johnson for the development of the slxfig module and the SED scripts that have been used to prepare the figures in this work.  

\item[Contributions]
The TANAMI program is coordinated by R.O. and M.K. 
F.K. led the multiwavelength data analysis and modeled the SED. 
K.M. led the theoretical interpretation of the SED data.
C.M., R.S., J.T., B.C., A.Ka1., E.R., R.O., and M.K. analyzed the LBA data. 
J.W. and N.G. were responsible for X-ray observations and data analysis.
T.B., S.B., C.G., C.M., D.E.G., A.Kr., K.L., E.L., F.L., T.S., and AZ contributed to the analysis and discussion of radio, optical/UV, X-ray and $\gamma$-ray data.
LBA observations were conducted by P.G.E., H.H., S.H., J.E.J.L., C.Ph., C.Pl., J.Q., J.S., and A.K.T. 
Hard X-ray data were  reduced and analyzed by T.B., I.Kr., W.B. and M.L.
G.A., T.E., D.El., C.J., A.Ka2., U.K., and M.Kr. contributed to the discussion of neutrino astronomy aspects.
M.K. and F.S. led the neutrino velocity discussion.    
D.J.T., R.O. and M.K. coordinated the TANAMI-LAT collaboration liaison. 
All authors discussed the results and commented on the manuscript.
  
 \item[Competing Interests] The authors declare that they have no
competing financial interests.
 \item[Correspondence] Correspondence and requests for materials
should be addressed to M.K. (email: \url{matthias.kadler@astro.uni-wuerzburg.de}).
\end{addendum}


\begin{methods}
\textbf{VLBI Imaging}\\
The TANAMI VLBI images of PKS~B1424$-$418 (Supplementary Data Fig.~\ref{fig:fig2}) are based
on observations on 2011 Nov 13, 2012 Sep 16, and 2013 Mar 14 with the Long
Baseline Array (LBA) and associated telescopes in Australia, New
Zealand, South Africa, Chile, and Antarctica. Correlation of the data
was performed with the DiFX correlator\cite{Del11} at Curtin
University in Perth, Western Australia. The data were calibrated and
imaged following Ojha~et~al.~(2010)\cite{Ojh10}. Supplementary Data Table \ref{tab:vlbi-images} gives an
overview of the array configuration and image parameters. 
The noise level for each image was determined by fitting a Gaussian function to an off-source part of the image. For this purpose and for producing the final images shown here we used the software package ISIS 1.6.2\cite{Hou00}.

The TANAMI array varies in configuration for each observation, leading to inhomogeneous sampling of the visibility data. This results in a different noise level and restoring beam in the final images. In order to compare the images, they were created with the maximum noise level of all three images (1.1\,mJy/beam) and the enclosing beam for convolution, i.e. the restoring beam which encompasses the synthesis beam of all observations (here $2.26 \times 0.79$\,mas$^2$ at a position angle of $9.5^\circ$).
The brightness temperature $T_\mathrm{B}$ and size of the central emission region diameter $d_\mathrm{core}$ from each observation were determined by fitting a two-dimensional Gaussian function to the visibility data. $T_\mathrm{B}$ was then calculated following standard procedures\cite{Kov05}.
By adopting a cosmology\cite{Spe07} with $H_0 = 73\mathrm{\,km\,s^{-1}\,Mpc^{-1}}$, $\Omega_\Lambda = 0.73$ and $\Omega_m = 0.27$ for the conversion from angular to linear scales, $1\mathrm{\,mas}$ corresponds to about $8.3\mathrm{\,pc}$.

\textbf{\textit{Fermi}/LAT $\gamma$-ray data analysis}\\
For the analysis of \textit{Fermi}/LAT $\gamma$-ray data, we used the \textit{Fermi} Science Tools (v9r32p5) with the
reprocessed Pass 7 data
and the P7REP\_SOURCE\_V15 instrument response functions\cite{Ack12} and a region of interest (ROI) of
10$^\circ$ 
in the energy range 100\,MeV to 300\,GeV. 
We applied a zenith angle limit of $100^\circ$ and the \texttt{gtmktime} cut \texttt{DATA\_QUAL==1 \&\& LAT\_CONFIG==1 \&\& ABS(ROCK\_ANGLE)<52}.
We used the Galactic diffuse emission model \texttt{gll\_iem\_v05.fits} and the isotropic diffuse model \texttt{iso\_source\_v05.txt}.
The model consists of all 2FGL sources inside the ROI. For the spectral and light-curve analysis, the normalization constants
of all model sources were left free. 
The source spectral index was left free for the spectral analysis and frozen for the light curve analysis during the modeling of the spectrum.
The $\gamma$-ray light curve of PKS~B1424$-$418 (Fig.~\ref{fig:fig1}) has
been calculated using 14 day bins.
We applied a Bayesian-blocks analysis\cite{Sca13} to the light curve. We used the
change points to determine the onset of the outburst as 2012 Jul 16, which is
marked as a shaded blue region in Fig.~\ref{fig:fig1}. 

All $\gamma$-ray sources detected by \textsl{Fermi}/LAT with flares
that exceeded the pre-defined threshold of
$10^{-6}\mathrm{cm^{-2}\,s^{-1}}$ ($>100$\,MeV) are included in the
\textit{LAT Monitored Source List}\footnote{Light curves of all these sources are
publicly available:~\url{http://fermi.gsfc.nasa.gov/ssc/data/access/lat/msl_lc/}.}. By
definition, this sample represents the brightest \textsl{Fermi}/LAT
sources. We used this list to identify the sources with the highest $\gamma$-ray fluences
and analyzed their $\gamma$-ray light curves in detail as described above.
We integrated the light curves from 2012 Jul 16 through 2013 Apr 30 
to determine the highest-fluence sources during the months around the arrival time of the 
2\,PeV IceCube neutrino event.
PKS~B1424$-$418 is the brightest blazar in terms of
$\gamma$-ray fluence: $F=(30.5\pm0.3)\,\mathrm{cm^{-2}}$ (the uncertainty is statistical only). The next on
the list are the blazars CTA\,102 ($F=(20.0\pm0.3)\,\mathrm{cm^{-2}}$), and B2~1633$+$38
($F=(18.2\pm0.3)\,\mathrm{cm^{-2}}$), which are 
both in the northern hemisphere.

\textbf{Broadband SED}\\
The broadband SED of PKS~B1424$-$418 (Fig.~\ref{fig:fig3}) has been built
from quasi-simultaneous data from 
\textit{Fermi}/LAT,
\textit{Swift}/XRT, \textit{Swift}/UVOT, SMARTS\cite{Bon12}, ALMA, ATCA and the LBA.
Archival data from 2MASS\cite{Sku06} and WISE\cite{Wri10}
catalogs are included as well. 
 
\textit{Swift}/XRT data were reduced with standard methods, using the most
recent software packages (HEASOFT 6.15.1) and calibration databases.
Spectra were grouped to a minimum signal-to-noise ratio of 5.
Spectral fitting was performed with ISIS 1.6.2\cite{Hou00}. We fitted the 
(0.5--10)\,keV energy band with an absorbed power law
model, which yielded good results for all time ranges.
The source showed no evidence of intrinsic X-ray absorption in excess
of the Galactic value\cite{Kal05}. 
X-ray data were deabsorbed using the best available abundances\cite{Wil00}
and cross-sections\cite{Ver96}. \textit{Swift}/UVOT data were
extracted following standard methods. 
Optical, infrared, and ultraviolet data were dereddened using
the same absorbing columns\cite{Nov12}.
At hard X-ray energies, we used data from \textit{Swift}/BAT and \textit{INTEGRAL}/IBIS from the energy bands of 20\,keV to 100\,keV and 20\,keV to 200\,keV, respectively. 
The source was not included in the BAT 70-month catalog\cite{Bau13}, which was based on data between 2004 Dec and 2010 Dec.
Using stacked BAT maps from the 104-month survey (2004 Dec through 2013 Aug), we detected the source and calculated the flux by applying a power-law fit to the spectrum.
For \textit{INTEGRAL}/IBIS we extracted a 3$\sigma$ upper limit from the variance of a mosaic combining all pointed observations in the time-range from 2011 Dec to 2013 Sep, where the source was located within 14\,$^\circ$ from the pointing direction.
The 104-month average BAT flux value is in good agreement with the 2LAC SED, whose data time period overlaps (2008 Aug to 2010 Aug). It is marginally consistent with the SEDs corresponding to the post-2010 short-flare and the three year IceCube period and well below the SED for the high-fluence period between 2012 Jul and 2013 Apr.  

The broadband spectra were fit with two logarithmic parabolas\cite{Mas04}
in order to parametrize the characteristic double humped
structure.
The spectral-model fit is performed in detector space while the fluxes in each energy bin in the top panel 
of Fig.~\ref{fig:fig3} are calculated independent of the model\cite{Now05}, under the assumption of constant flux in each bin. Note that for curved spectra and wide energy bins, this approach can lead to an apparent positive offset of the flux data points over the best-fit model as seen, e.g., in the \textit{Fermi}/LAT data points of the `short flare' in Fig.~\ref{fig:fig3}.
X-ray absorption and optical extinction were included in
the broadband fit. Upper limits have generally not been considered in the
fits.
The low $\gamma$-ray upper limits at the highest \textit{Fermi}/LAT energies indicate a possible cutoff.
Including the upper limits as values in the SED fits forces the parabolas to increase in peak flux in order to still describe the slope at X-ray energies and reduces the overall fit quality. The increase in peak flux leads to greater numbers of neutrinos (see below), with the exception of the short flare, so that our method to ignore the upper limits in the fit is more conservative for our purposes.

\textbf{Neutrino event prediction}\\
To derive maximum neutrino events to be expected from the measured SEDs,
we applied the techniques that we developed for our analysis of the AGN in the fields of the first two PeV IceCube events. 
We used 2FGL GeV $\gamma$-ray spectra\cite{2fgl} and extracted X-ray spectra taken during the same period by \textit{Swift}\cite{Geh04}. For sources not observed by \textit{Swift}, we used X-ray data or upper limits from the \textit{ROSAT} all-sky survey\cite{rass}.
For PKS~B1424$-$418, we calculated in addition \textit{Fermi}/LAT $\gamma$-ray spectra for the various time ranges of interest according to the methods described above.
We fit a single log parabola to the high-energy data.
From the integrated 1\,keV to 5\,GeV SEDs, we can derive a maximum-possible neutrino flux for each source. From this we can derive the maximum-possible number of counts for three years of IceCube integration (see Supplementary Data Table~\ref{tab:events}) using the appropriate effective area for northern/southern-hemisphere sources for electron neutrinos of the IceCube 998-day analysis\cite{Ice14}.
Formal uncertainties of neutrino counts can be derived from the uncertainties of the integrated electromagnetic energy output but are dominated by systematic uncertainties due to the simplicity of the model.

\textbf{Calculation of chance coincidences of major blazar outbursts and PeV neutrino events}\\
We estimate the number of chance coincidences of major blazar outbursts and PeV neutrino events \textsl{a posteriori} using $N_\mathrm{coinc}= \dot{N_{\nu}} \times \dot{N}_{\textrm{outburst}} \times \tau \times T$. Here, $\dot{N_{\nu}} $ is the number of PeV-neutrinos per year in the southern sky, $\Omega_\mathrm{south} = 2\pi$, $\dot{N}_{\mathrm{outburst}}$ is the number of major outbursts per year within a cone covered by the typical median uncertainty of shower-like PeV neutrino events  of $\Omega_\nu \sim 0.16$\,sr (averaged median angular uncertainty of the three observed PeV events $\sim 13^\circ$), $\tau$ is the coincidence time window defined by the outburst time, and $T$ is the IceCube observation time of $T=3$\,yr, during which IceCube has observed three PeV events. We therefore obtain for $\dot{N_{\nu}} \sim 1\,\mathrm{yr}^{-1}$. 
High-fluence blazar outbursts occur on time scales of months so we choose, conservatively, $\tau = 1$\,yr for the time window for chance coincidences.
In the \textit{Fermi}/LAT blazar light curves for six years and for the whole sky, we find 
a total of eight 1\,yr periods, during which an individual 
blazar reached or exceeded the GeV fluence of the PKS~B1424$-$418 outburst.
These high-fluence phases occured in the sources 
PKS~B1424$-$418, PKS~B\,1222$+$216, 
3C\,273, 3C\,454.3 (2 periods), and PKS\,B1510$-$089 (3 periods).
Hence, the rate of such outbursts occurring within $\Omega_\nu$  is expected to be $\dot{N}_{\mathrm{outburst}} \sim 0.1\,\mathrm{yr}^{-1}\,\mathrm{sr}^{-1} \times \Omega_\nu = 0.016\,\mathrm{yr}^{-1}$. Using these numbers we end up with $N_\mathrm{coinc} \sim 0.05$ as the mean number of chance coincidences. The Poisson probability to observe one or more such coincidences is about 5\%. However, because of the lack of a pre-defined statistical test, we cannot formally use this value to test the hypothesis of a chance coincidence. 

\textbf{Coincidences of sub-PeV neutrino events with high-fluence blazars}\\
All high-fluence blazars from Table~\ref{tab:rankedlist} are in coincidence with more than one IceCube sub-PeV neutrino event, which is not surprising given the high probability of chance coincidences within the large $R_{50}$ radii of the cascade-like IceCube events. In addition, up to 25 of the 34 sub-PeV events are of possible atmospheric origin\cite{Ice14}.  The 
most interesting cases are IC\,7, which coincides with
long-lasting high-fluence states of the two high-fluence blazars PKS~B0537$-$441 and PKS~B2326$-$502, which are both located within the $R_{50}$ region of IC\,7, similar to the one observed in PKS~B1424$-$418 and coinciding with the 2\,PeV event IC\,35. The integrated fluence of PKS~B0537$-$441 and PKS~B2326$-$502 predicts a maximum of $\sim5$ neutrino events, i.e., applying the scaling factor of 0.025, the Poisson probability for the detection of more than zero PeV events is about 12\,\%. 
The IceCube events IC\,16 and IC\,25 are close ($1.9 R_{50}$ and $1.2 R_{50}$, respectively) to 
the position of the top-ranked high-fluence source, PKS~B1510$-$089. 
It is remarkable that the two observed events occurred close to the highest peaks of the GeV light curve  but we emphasize again that a considerable fraction of IceCube events at these comparatively low energies can be of atmospheric origin and that the field of IC\,25 is particularly large. 

\textbf{Coincidences of track neutrino events with $\gamma$-ray blazars}\\
We note that the IceCube track events, which have a much better angular resolution than the shower events, are 
all of rather low energy and/or are located at  far northern declinations. 
Most of them are thus likely of atmospheric origin.
In fact, event IC\,28 showed associated hits in the IceTop surface air shower array\cite{Ice13,Ice14}. The IceCube team considers three additional track events particularly ambiguous, which started near the detector boundary and were down going (IC\,3, IC\,8, IC\,18, IC\,23). The positions of the remaining three (IC\,5, IC\,13, and IC\,37) which can be considered the most likely track events to be of true extraterrestrial origin, have been compared with positions of known $\gamma$-ray AGN. One of these events (IC\,5) is spatially coincident with the blazar PKS\,0723$-$008, but as pointed out previously\cite{Bro15}, the chance probability for a \textit{Fermi}/LAT detected AGN to coincide with any given field of about 1\,deg$^2$ is high. We estimate the Poisson probability for IC\,5 from the known 1017 $\gamma$-ray AGN in the 2\,LAC catalog\cite{2lac} distributed over the full sky (excluding the $|b|<10^\circ$ region around the Galactic plane) to be 41\,\%. However, as mentioned already above, only half of all IceCube events are expected to hold the coordinates of their true astrophysical sources inside their $R_{50}$ regions, while the other half are located at larger positional offsets. In this small sample of three track events, we thus expect only one or two coincidences within $R_{50}$. If we consider radii twice as large, we find one 2\,LAC blazar for IC\,13 (4C\,+41.11 at $1.8 R_{50}$) and one for IC\,37 (RBS\,0958 at $1.6 R_{50}$). The \textsl{a posteriori} chance probability for all three track events to coincide within $2 R_{50}$ with a 2\,LAC blazar is 22\,\%. None of the candidate blazars is of particularly high fluence. It is likely that they represent the large population of `typical' 2LAC $\gamma$-ray blazars, which have low individual detection probabilities. However, their integrated emission plus the contribution of fainter unresolved blazars to the EGB is high and may lead to the detected events.

\end{methods}

\clearpage

\clearpage

\clearpage

\begin{figure}
\centering
\includegraphics[height=0.8\textheight]{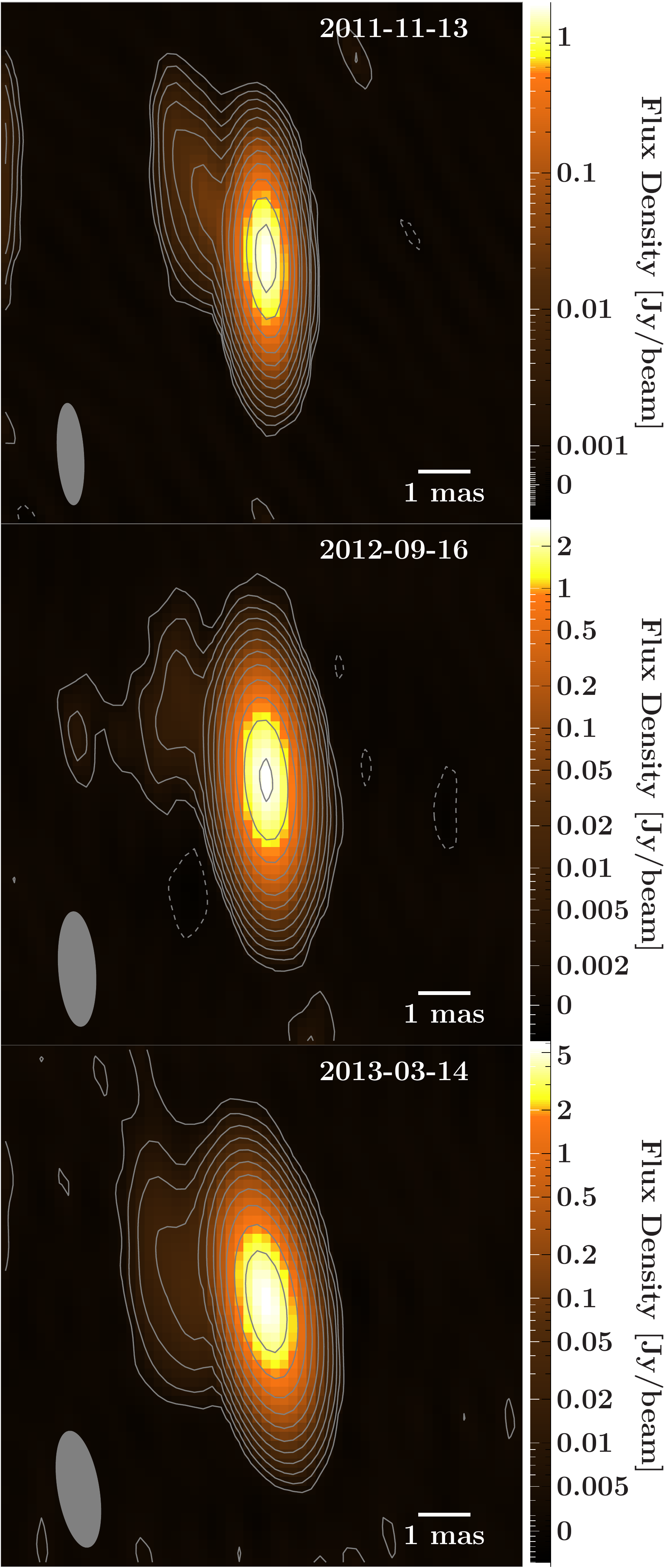} %
\caption{\small \textbf{Supplementary Data Figure 1} -- Individual TANAMI VLBI images of the core region of PKS~B1424$-$418. Each image corresponds to the properties in Supplementary Data Table \ref{tab:vlbi-images} for the individual observations. All contours start at $3 \sigma_\textrm{RMS}$ and increase logarithmically by factors of 2 .  \label{fig:fig4} 
}
\end{figure}

\clearpage

\begin{table}
\caption{\textbf{Maximum-possible number of PeV neutrino events in 988 days of IceCube data for the 17 2LAC $\gamma$-ray blazars in the field of the 2\,PeV IceCube event based on 2LAC catalog $\gamma$-ray spectra and contemporaneous X-ray data. 
}}
\vspace*{\baselineskip}
\footnotesize 

\label{tab:events}
\centering
\begin{tabular}{@{}llrrr@{}}
\hline 	
2FGL Name & Common Name & $F_\gamma($erg~cm$^{-2}$s$^{-1}$) &  $N_{\nu,\mathrm{PeV}}^\mathrm{max}$ &   \\
\hline
2FGL J1230.2$-$5258	&	PMN J1229$-$5303	&	$\left(2.4^{+1.5}_{-1.5}\right)\times10^{-11}$	&	0.14	\\
2FGL J1234.0$-$5733	&	PMN J1234$-$5736	&	$\left(1.1^{+0.4}_{-0.4}\right)\times10^{-11}$	&	0.06	\\
2FGL J1303.5$-$4622	&	PMN J1303$-$4621	&	$\left(1.9^{+0.6}_{-0.6}\right)\times10^{-11}$	&	0.11	\\
2FGL J1303.8$-$5537	&	PMN J1303$-$5540	&      $\left(1.04^{+0.11}_{-0.11}\right)\times10^{-10}$&	0.38	\\
2FGL J1304.3$-$4353	&	1RXS 130421.2$-$435308	&	$\left(2.11^{+0.25}_{-0.25}\right)\times10^{-11}$&	0.12	\\
2FGL J1307.5$-$4300	&	1RXS 130737.8$-$425940	&	$\left(8.4^{+1.7}_{-1.7}\right)\times10^{-12}$	&	0.05	\\
2FGL J1307.6$-$6704	&	PKS~B 1304$-$668	&	$\left(1.54^{+0.15}_{-0.15}\right)\times10^{-10}$&	0.89	\\
2FGL J1314.5$-$5330	&	PMN J1315$-$5334	&	$\left(8.1^{+0.9}_{-0.9}\right)\times10^{-11}$	&	0.47	\\
2FGL J1326.7$-$5254	&	PMN J1326$-$5256	&	$\left(1.04^{+0.21}_{-0.18}\right)\times10^{-10}$&	0.59	\\
2FGL J1329.2$-$5608	&	PMN J1329$-$5608	&	$\left(1.38^{+0.36}_{-0.29}\right)\times10^{-10}$&	0.93	\\
2FGL J1330.1$-$7002	&	PKS~B 1326$-$697	&	$\left(1.53^{+0.11}_{-0.11}\right)\times10^{-10}$&	0.89	\\
2FGL J1352.6$-$4413	&	PKS~B 1349$-$439	&	$\left(5.4^{+1.0}_{-1.0}\right)\times10^{-11}$	&	0.32	\\
2FGL J1400.6$-$5601	&	PMN J1400$-$5605	&	$\left(6.9^{+0.8}_{-0.8}\right)\times10^{-11}$	&	0.40	\\
2FGL J1407.5$-$4257	&	CGRaBS J1407$-$4302	&	$\left(1.6^{+0.5}_{-0.5}\right)\times10^{-11}$	&	0.09	\\
2FGL J1428.0$-$4206	&	PKS~B1424$-$418	&	$\left(2.04^{+0.17}_{-0.16}\right)\times10^{-10}$&		1.57	\\
2FGL J1508.5$-$4957	&	PMN J1508$-$4953	&	$\left(7.6^{+3.0}_{-2.3}\right)\times10^{-11}$&		0.55	\\
2FGL J1514.6$-$4751	&	PMN J1514$-$4748	&	$\left(5.6^{+0.6}_{-0.6}\right)\times10^{-11}$	&	0.32	\\
\hline
Sum (2LAC)              &                               &                                                       &       7.9    \\
\end{tabular}
\end{table}

\begin{table}
\caption{\textbf{Ranked list of the 10 highest-fluence blazars during the first three years of IceCube observations. Note that the IceCube effective area is substantially smaller for northern-hemisphere PeV sources than for the southern sky.  
}}
\vspace*{\baselineskip}
\footnotesize
\label{tab:rankedlist}
\centering\footnotesize
\begin{tabular}{@{}lrrccccc@{}}
\hline 	
Name & R.A. & Dec. & $F_\gamma$&  $N_{\nu,\mathrm{PeV}}^\mathrm{max}$ & $N_{\nu,\mathrm{PeV}}^\mathrm{pred}$ & $N_{\nu,\mathrm{PeV}}^\mathrm{pos}$ \\
     & [Deg]& [Deg]&    [$10^{-10}$erg~cm$^{-2}$s$^{-1}$]          &    & &        \\ 
\hline
PKS~B1830-21   & 278.4 & $-$21.1 & $\left(14.34\pm0.27\right)   $& 8.3 & 0.21 & 1 & \\
PKS~B1510$-$089 & 228.2 & $-$9.1  & $\left(13.31\pm0.13\right)$& 7.7 & 0.19 &  0 &    \\
3C\,454.3     & 343.5 & 16.2    & $\left(37.50\pm01.3\right)  $& 7.6 & 0.19 &  0 & \\
PKS~B1424$-$418 & 217.0 & $-$42.1 & $\left(7.82\pm0.16\right) $& 4.5 & 0.11 &  1 &\\
PKS~B2326$-$502 & 352.3 & $-$49.9 & $\left(4.69\pm0.10\right) $& 2.7 & 0.07 &  0 &\\
PKS~B0537$-$441 & 84.7  & $-$44.1 & $\left(3.84\pm0.08\right) $& 2.2 & 0.06 &  0 &\\
PKS~B1222$+$216&186.2 & 21.4    & $\left(7.94\pm0.12\right)   $& 1.6 & 0.04 &  0 &\\
CTA\,102      & 338.2 & 11.7    & $\left(6.42\pm0.12\right)   $& 1.3 & 0.03 &  0 &\\
B2~1633$+$38& 248.8& 38.1 & $\left(6.28\pm0.09\right)   $& 1.3 & 0.03 &  0 &\\
B2~1520+31    & 230.5 & 31.7    & $\left(4.75\pm0.25\right)   $& 1.3 & 0.02 &  0 &\\
\hline
\end{tabular}
\begin{minipage}{\linewidth}\footnotesize
Note: Columns (1): Source name; (2,3): J2000 coordinates; (4): integrated $\gamma$-ray flux between 5\,keV and 10\,GeV; (5,6): maximum-possible and predicted number of neutrino detections; (7): number of PeV events that might be associated with each source based on a positional coincidence.   
\end{minipage}
\end{table}

\begin{table}
\caption{\textbf{Details of TANAMI VLBI observations of PKS~B1424$-$418 and image parameters}}
\label{tab:vlbi-images}
\centering
\begin{tabular}{@{}l@{}cccc@{}c@{}ccc@{}}
\hline 	
Date & $S_\mathrm{total}$ & $S_\mathrm{peak}$ & $\sigma_\mathrm{RMS}$ &  $b_\mathrm{maj}$ & $b_\mathrm{min}$ & P.A. & $T_\mathrm{B,core}$ & $d_\mathrm{core}$\\
{}[yyyy-mm-dd] & [Jy] & [Jy/beam] & [mJy/beam] & [mas] & [mas] & [$^\circ$] & [$10^{11}\mathrm{\,K}$] & [pc]\\
\hline
2011 Nov 13 & $2.33\pm 0.23$ & $1.80\pm 0.18$ & $0.2$ & $1.95$ & $0.50$ & $3.8$ & $4.2$ & $2.5$\\
2012 Sep 16 & $3.17\pm0.32$ & $2.89\pm0.29$ & $0.4$ & $2.20$ & $0.71$ & $4.0$ & $7.4$ & $2.2$\\
2013 Mar 14 & $6.23\pm0.62$ & $5.61\pm 0.56$ & $1.1$ & $2.26$ & $0.79$ & $9.5$ & $13$ & $2.3$\\
\hline
\end{tabular}
\begin{minipage}{\linewidth}\footnotesize
Note: Tabulated image parameters refer to the formal values after hybrid imaging using natural weighting\cite{Ojh10}. The three images shown in Fig.~\ref{fig:fig3} are all convolved with the same restoring beam and use uniform contour lines and color scaling. See Fig.~\ref{fig:fig4} for the individual images restored with the individual beams and contours starting at $3 \sigma_\textrm{RMS}$. Core FWHM diameters $d_\mathrm{core}$ and $T_\mathrm{B,core}$ have been derived from fitting an elliptical Gaussian component to the visibility data.\\
Array configurations: PK-AT-HO-CD-TC-TD-HH-MP-WW (2011 Nov 13); PK-AT-HO-CD-TC-HH-AK-KE (2012 Sep 16); PK-AT-HO-CD-TC-TI-HH-WW-AK-KE (2013 Mar 14); PK: Parkes (64\,m), AT: ATCA (5x22\,m), HO: Hobart (26\,m), CD: Ceduna (30\,m), TC: TIGO (6\,m), TI: Tidbinbilla (70\,m), TD: Tidbinbilla (34\,m), MP: Mopra (22\,m), HH: Hartebeesthoek (26\,m), WW: Warkworth (12\,m), AK: ASKAP (12\,m), KE: Katherine (12\,m).
\end{minipage}
\end{table}

\end{document}